\documentclass[conference]{IEEEtran}
\IEEEoverridecommandlockouts
\usepackage{cite}
\usepackage{amsmath,amssymb,amsfonts}
\usepackage{algorithmic}
\usepackage{graphicx}
\usepackage{textcomp}
\usepackage{xcolor}
\usepackage{hyperref}

\def\BibTeX{{\rm B\kern-.05em{\sc i\kern-.025em b}\kern-.08em
    T\kern-.1667em\lower.7ex\hbox{E}\kern-.125emX}}
\begin{document}
\title{Deep Learning-Based Automated Segmentation of Uterine Myomas}
\author{\IEEEauthorblockN{Tausifa Jan Saleem}
\IEEEauthorblockA{\textit{Department of Computer Vision} \\
\textit{Mohamed Bin Zayed University of Artificial Intelligence}\\
UAE}\\
\and
\IEEEauthorblockN{Mohammad Yaqub}
\IEEEauthorblockA{\textit{Department of Computer Vision} \\
\textit{Mohamed Bin Zayed University of Artificial Intelligence}\\
UAE} \\
}
\maketitle
\section{Introduction}
Uterine fibroids (myomas) are the most common benign tumors of the female reproductive system, particularly among women of childbearing age. With a prevalence exceeding 70\%, they pose a significant burden on female reproductive health \cite{mcwilliams2017recent}. Clinical symptoms such as abnormal uterine bleeding, infertility, pelvic pain, and pressure-related discomfort play a crucial role in guiding treatment decisions, which are largely influenced by the size, number, and anatomical location of the fibroids \cite{mcwilliams2017recent}. Magnetic Resonance Imaging (MRI) is a non-invasive and highly accurate imaging modality commonly used by clinicians for the diagnosis of uterine fibroids. Segmenting uterine fibroids requires a precise assessment of both the uterus and fibroids on MRI scans, including measurements of volume, shape, and spatial location. However, this process is labor intensive and time consuming and subjected to variability due to intra- and inter-expert differences at both pre- and post-treatment stages. As a result, there is a critical need for an accurate and automated segmentation method for uterine fibroids. \par In recent years, deep learning algorithms have shown remarkable improvements in medical image segmentation, outperforming traditional methods. These approaches offer the potential for fully automated segmentation. Several studies have explored the use of deep learning models to achieve automated segmentation of uterine fibroids \cite{tinelli2025artificial}. However, most of the previous work has been conducted using private datasets, which poses challenges for validation and comparison between studies \cite{tinelli2025artificial}. In this study, we leverage the publicly available Uterine Myoma MRI Dataset (UMD)\cite{pan2024large} to establish a baseline for automated segmentation of uterine fibroids, enabling standardized evaluation and facilitating future research in this domain.
\section{Materials and methods} \label{sec:methods}
\subsection{Dataset}
\begin{figure}[h]
\centering
\includegraphics[scale=0.5]{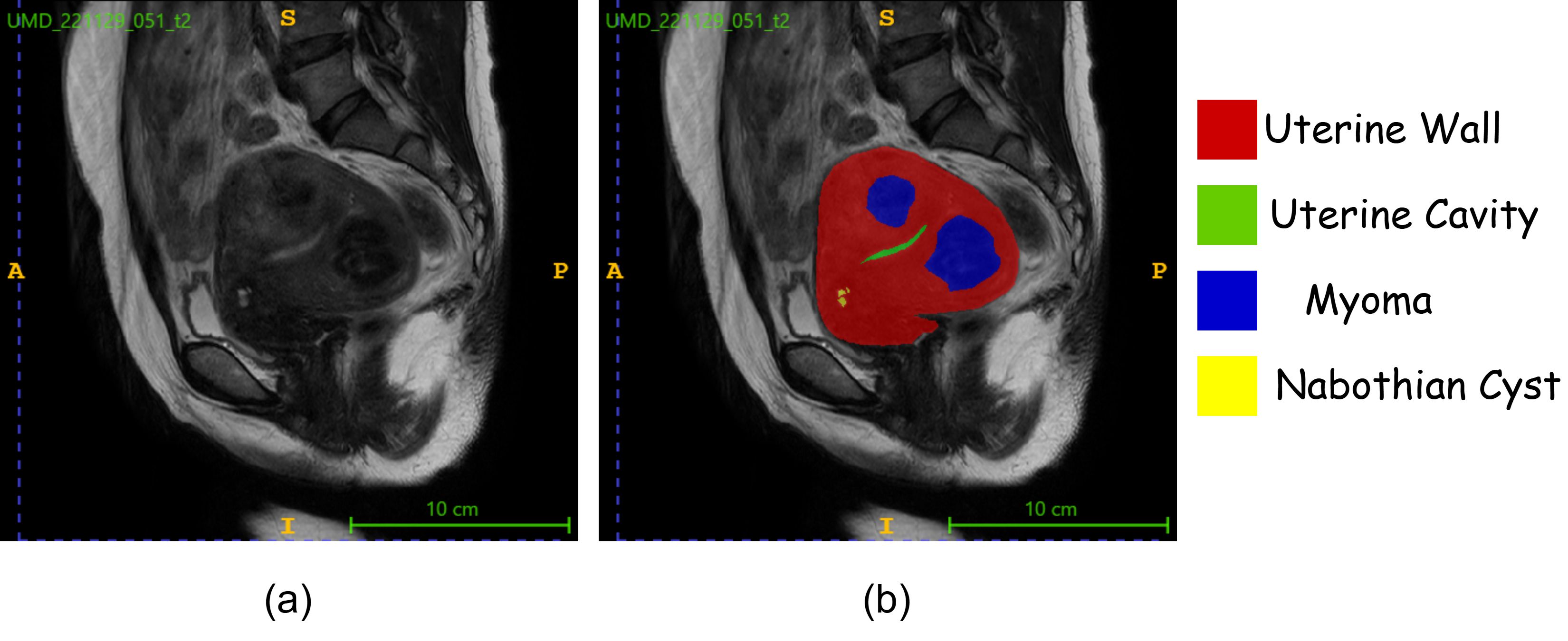}
\caption{UMD data example (a) Sagittal MRI slice, (b) Sagittal MRI slice with overlaid segmentation labels.}
\label{fig1}
\end{figure}

UMD \cite{pan2024large} is a publicly available dataset comprising sagittal T2-weighted pelvic MRI scans from 300 patients diagnosed with uterine myoma. It provides pixel-level annotations of four structures: uterine wall, uterine cavity, myoma, and nabothian cyst. Collected between 2015 and 2023, the dataset includes patients aged 21 to 86 years (mean±SD: 49.73±12.96), and serves as a high-quality, expert-validated resource for developing and evaluating medical image segmentation algorithms. Figure \ref{fig1} presents a representative example from the UMD dataset, showing a sagittal T2-weighted MRI slice and the same slice with overlaid segmentation labels.
\subsection{Methodology}
We propose a method based on the nnU-Netv2 framework \cite{isensee2021nnu} for fully automated segmentation of the uterine wall, uterine cavity, uterine myoma, and nabothian cyst using the UMD dataset. MRI scans from 246 patients were used for training, and the remaining 54 scans were reserved for testing. The proposed method uses a U-Net-based encoder-decoder architecture with five resolution levels, where each level integrates convolutional blocks with instance normalization and leaky ReLU activation. Downsampling is performed via strided convolutions, while upsampling relies on transposed convolutions to restore spatial resolution. No manual hyperparameter tuning was required, as nnU-Netv2 automatically configures its architecture and training parameters based on the dataset characteristics. \par We utilized the 3D full-resolution variant and trained the model for 400 epochs using a composite loss function combining Dice loss and cross-entropy loss to jointly optimize region overlap and voxel-wise classification performance. The total loss $\mathcal{L}_{\text{total}}$ is defined as:
\begin{equation}
\mathcal{L}_{\text{total}} = \mathcal{L}_{\text{Dice}} + \mathcal{L}_{\text{CE}}.
\end{equation}
The multi-class Dice loss $\mathcal{L}_{\text{Dice}}$ is given by:
\begin{equation}
\mathcal{L}_{\text{Dice}} = 1 - \frac{2 \sum_{c=1}^{C} \sum_{i=1}^{N} p_{i,c} g_{i,c}}{\sum_{c=1}^{C} \sum_{i=1}^{N} p_{i,c}^2 + \sum_{c=1}^{C} \sum_{i=1}^{N} g_{i,c}^2 + \epsilon},
\end{equation}
where $C$ is the number of classes, $N$ is the total number of voxels, $p_{i,c}$ is the predicted probability for voxel $i$ and class $c$, $g_{i,c}$ is the ground truth for voxel $i$ and class $c$, $\epsilon$ is a small constant to avoid division by zero.
The cross-entropy loss $\mathcal{L}_{\text{CE}}$ is computed as:
\begin{equation}
\mathcal{L}_{\text{CE}} = - \sum_{i=1}^{N} \sum_{c=1}^{C} g_{i,c} \log(p_{i,c}).
\end{equation}
The performance of the model was evaluated using the Dice similarity coefficient (DSC), also referred to as the Dice score, which measures the spatial overlap between the predicted and ground truth segmentations. \par
\begin{figure}[h]
\centering
\includegraphics[scale=0.41]{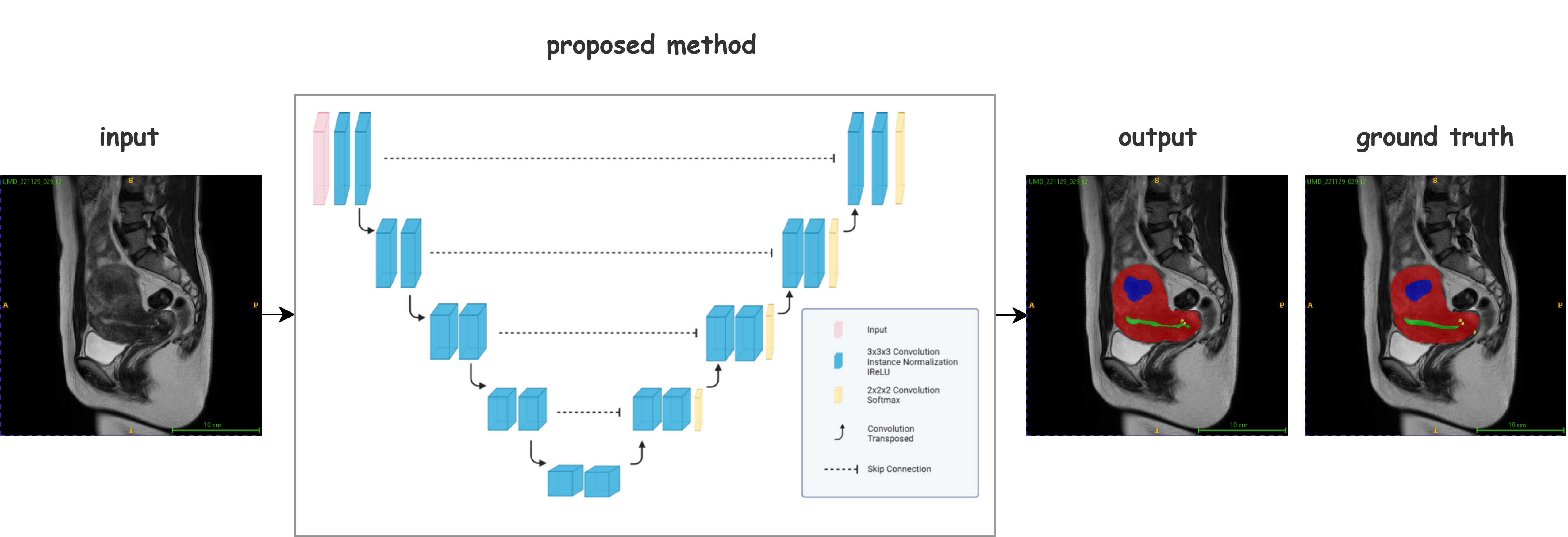}
\caption{Uterine MRI segmentation during inference on the test data.}
\label{fig2}
\end{figure}

 An example of the segmentation output during inference on the test set using the proposed method is shown in Figure~\ref{fig2}. The close alignment between the predicted segmentation and the ground truth on the test set example highlights the effectiveness of the method in automated uterine MRI segmentation.
\section{Results and discussions}
\begin{figure}[h]
\centering
\includegraphics[scale=0.3]{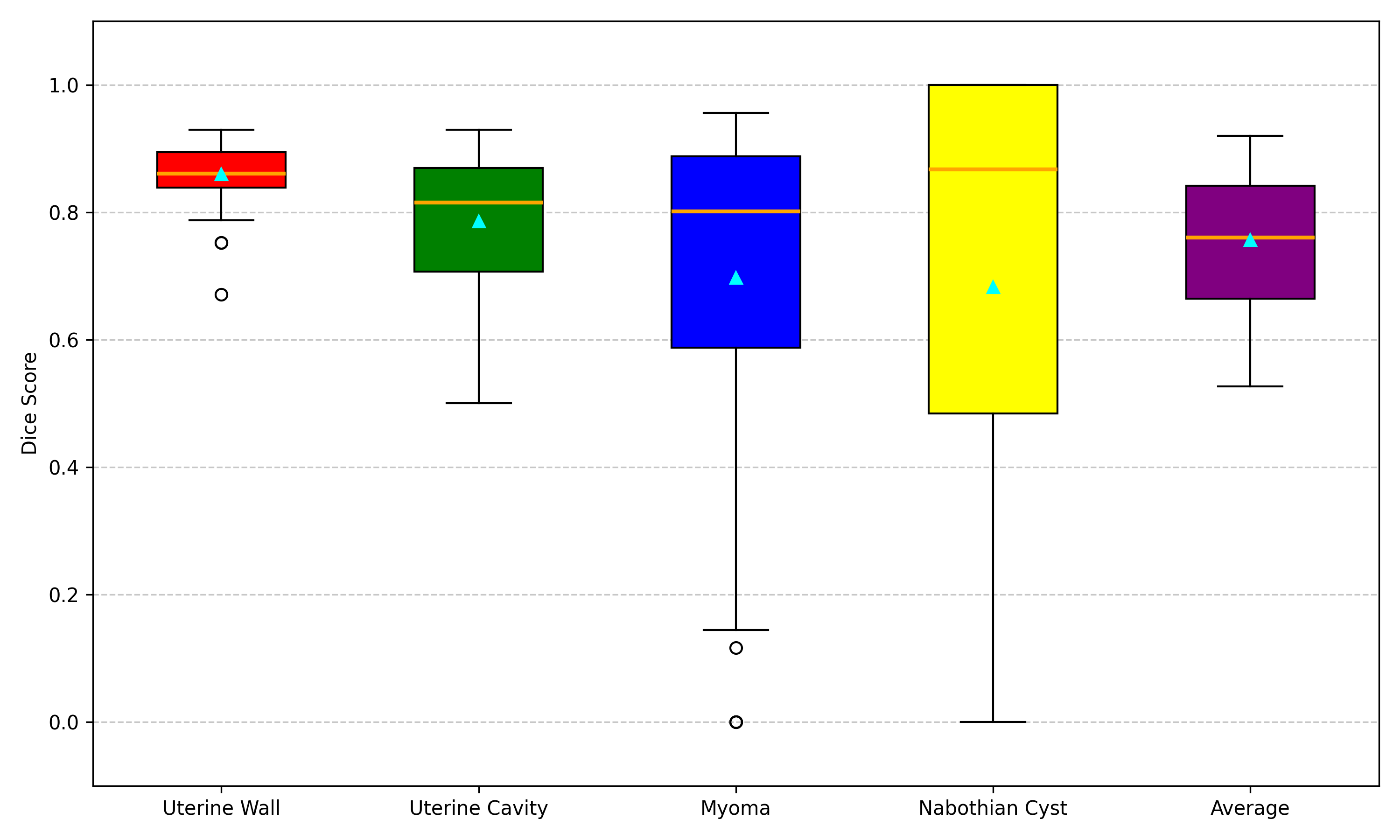}
\caption{Dice Scores for uterine wall, uterine cavity, myoma and nabothian cyst on the test set.}
\label{fig3}
\end{figure}

Despite the inherent challenges posed by myomas, including their variable shape, size, and appearance, the proposed method achieved a mean Dice score of 0.70 for this class (Table \ref{tab:example}), highlighting its robustness in detecting and delineating myomas across diverse cases. The large standard deviation reflects substantial case-to-case variability, as also visualized in Figure \ref{fig3} where some cases exhibit Dice scores below 0.50, while some others exceed 0.80. This suggests that while the model can segment myomas accurately in favorable cases, challenges remain for cases with atypical morphology or low contrast. Such variability underscores the importance of further refinements, such as tailored post-processing or additional training data focused on difficult cases.\par Beyond myoma segmentation, the proposed method also performed impressively across other classes. It achieved Dice scores of 0.86 for the uterine wall, 0.79 for the uterine cavity, and 0.68 for the nabothian cyst (Table \ref{tab:example}), demonstrating the method's ability to generalize across multiple classes. Notably, uterine wall segmentation was highly consistent across cases, as reflected by its low standard deviation and compact box plot distribution (Table \ref{tab:example}, Figure \ref{fig3}). In contrast, the lower and more variable scores for the nabothian cyst reflect expected challenges associated with its small size. Overall, the proposed method achieved a mean Dice score of 0.76 across all classes, validating its effectiveness as a powerful framework for uterine MRI segmentation. Crucially, its performance on myoma segmentation establishes a significant step forward in the development of automated tools for managing one of the most prevalent and burdensome conditions in women's reproductive health.
\begin{table}[h]
\centering
\caption{Dice scores (mean ± standard deviation) on the test set.}
\label{tab:example}
\begin{tabular}{c|c}
\hline
\textbf{Label} & \textbf{Dice Score} \\ \hline
Uterine Wall & 0.86 ± 0.05  \\ 
Uterine Cavity & 0.79 ± 0.10 \\ 
Myoma & 0.70 ± 0.27  \\ 
Nabothian Cyst & 0.68 ± 0.38  \\ \hline
\end{tabular}
\end{table}
\section{Conclusions}
In this study, we addressed the need for an automated and reproducible method for uterine myoma segmentation by leveraging a publicly available MRI dataset. The results demonstrate that our proposed method can reliably segment uterine myomas along with surrounding structures, offering a robust baseline for future research. This approach holds promise for reducing the clinical burden associated with manual segmentation and enabling more standardizeded assessment of myoma characteristics.

\end{document}